\documentclass[aps,prd,12pt,oneside,tightenlines,secnumarabic,a4paper,eqsecnum,nofootinbib]{revtex4}
\usepackage{amsmath,amssymb,amsfonts,latexsym,graphicx,fancyhdr}
\begin{document}

\title{Comment on ``Measurement of quantum states of
neutrons in the Earth's gravitational field"}

\author{Johan \surname{Hansson}}
\email[Corresponding author. Email address: ]{hansson@mt.luth.se}
\author{David \surname{Olevik}}
\author{Christian \surname{T\"{u}rk}}
\email[Email address: ]{chrtur-8@student.luth.se}
\author{Hanna \surname{Wiklund}}
\affiliation{Department of Physics \\ Lule{\aa} University of
Technology \\ SE-971 87 Lule\aa, Sweden}

\date{\today}

\begin{abstract}
In the paper by V.V. Nesvizhevsky \textit{et al.}, Phys. Rev. D
\textbf{67}, 102002 (2003), it is argued that the lowest quantum
state of neutrons in the Earth's gravitational field has been
experimentally identified. While this is most likely correct, it
is imperative to investigate all alternative explanations of the
result in order to close all loopholes, as it is the first
experiment ever claimed to have observed gravitational quantum
states. Here we show that geometrical effects in the experimental
setup can mimic the results attributed to gravity. Modifications
of the experimental setup to close these possible loopholes are
suggested.
\end{abstract}


\maketitle

\section{Introduction}
A well known property of quantum mechanics is the quantization of
the energy levels of a confined particle, \textit{e.g.}, one
trapped in a potential well. For instance, the electromagnetic and
the strong nuclear forces create different kinds of quantized
structure in atoms and nuclei. This suggests that a splitting of
the energy levels should also be observed for particles in the
Earth's gravitational field, but since the gravitational
interaction is much weaker, the effect is subtle and hard to
detect.

Recently \cite{Nev}, Nesvizhevsky \textit{et al.}, described an
experiment where such quantum effects of gravity acting on
ultra-cold neutrons (UCN) were claimed to have been observed. The
results of the experiment were also previously summarized in
\cite{Nev1}. UCN were allowed to flow through a cavity with a
reflecting surface below and an absorber above. By measuring the
number of neutrons exiting the experimental setup, they claim to
have observed discrete gravitational energy levels. They argue
that the discrete data is related to the sudden increase of
neutrons coming through at distinct widths between the reflecting
surface and absorber. However, since the UCN are restricted by
both the reflecting surface and the (non-ideal) absorber, also the
geometric effect should be considered. The claimed result may even
be explained by the use of geometrical arguments only.

\section{The experiment}
Here we give a brief review of the experiment reported in
\cite{Nev,Nev1}. A similar experiment was first suggested by V.I.
Luschikov and A.I. Frank in 1978 \cite{Qeffect78}.

The absorber and the ``mirror'' create a slit through which the
neutrons pass, eventually reaching a detector at the end of the
experimental setup. UCN are essential to the experiment due to
their crucial properties. First and foremost they are electrically
neutral, making them insensitive to ``stray" electric fields which
could easily mask all gravitational effects. They also have an
energy of about $10^{-7} \ \mathrm{eV}$, corresponding to a de
Broglie wavelength of $\sim $500 \AA \ or a (horizontal) velocity
of $\sim $10 m/s, allowing them to undergo total reflection at all
angles against a number of materials. The low energy also allows
for high resolution, and since neutrons have a lifetime of the
order of 900 s, it is possible to store them for periods of 100 s
or more.

Nesvizhevsky \textit{et al.}, argue that when the neutrons are
trapped in the potential formed by the mirror (an impenetrable
``floor") and the Earth's effectively linear gravitational
potential there will be a discrete set of possible energy levels,
$E_n$, corresponding to the allowed eigenfunctions $\psi_n$. These
are related through the time-independent Schr\"{o}dinger equation,
$H \psi_n = E_n \psi_n$, where $H = p^2/2m + V$, and $V = mgz$.
For a theoretical treatment of this potential, see \cite{PracQ}.
The four lowest theoretical energy eigenvalues are $E_1 = 1.4 \
\mathrm{peV}$, $E_2 = 2.5 \ \mathrm{peV}$, $E_3 = 3.3 \
\mathrm{peV}$ and $E_4 = 4.1 \ \mathrm{peV}$.

The ground state energy $E_1$ corresponds to a classical height,
$E_1=mgz$, of about 15 $\mu$m. This leads the group to predict
that when the slit-opening is less than this height no neutron
transmission will occur. They argue that if the quantum mechanical
wave function has a spatial extension larger than the opening, it
will not ``fit" without overlapping the absorber, and the neutrons
have no chance of reaching the detector. In the experiment they
observed a discrete increase in the number of detected neutrons as
the slit opening was increased. In particular it was observed, as
predicted, that when the slit-opening was less than $\sim$ 15
$\mu$m no neutrons reached the detector, and that there occurred a
sudden increase after 15 $\mu$m.

\section{Alternative explanations}
A first thing to emphasize is that the energy eigenvalues
themselves never were measured, \textit{i.e.}, all quoted energies
are entirely theoretical. The only experimental data are the
neutron counts $N$ at the detector, as a function of the
mirror-absorber slit-width $\Delta h$. The experimental statistics
for discrete steps corresponding to excited quantum levels is
insufficient \cite{Nev,Nev1}. Hence, the authors claim only to
conclusively have identified the quantized ground state (first
step). There is thus no absolute need to recreate the quoted
energy eigenvalues, as one only needs to explain the first jump in
the number of detected neutrons. The data also show a good fit to
a ``translated" classical curve $N \propto (\Delta h - h_1)^{1.5}$
(dotted curve in Fig. 5(c) of \cite{Nev}) in which only the first
discrete step is taken into account.

However, to show that it can be done, we choose as a first rough
approximation a potential consisting of two infinite walls,
\textit{i.e.}, a ``neutron in an infinite box". (The mirror can be
seen as an ``almost" infinite wall but the absorber is obviously
poorly described by this.) The mirror introduces a
non-gravitational ``external" force to obtain the confining
potential. This is very different from \textit{e.g.}, the Hydrogen
atom where the ``mirror" acting on the electron is internal,
arising from electromagnetic interaction and quantum uncertainty
only. The problem is trivial to solve analytically
\cite{PracQ,GroundQM}, and the allowed energies are
\begin{equation} \label{eq: Energy eigenvalues partic in box}
  E_n^{\mathrm{Box}} = \frac{\hbar^2\pi^2n^2}{2ma^2},
\end{equation}
where $a$ is the box-width. Thus, the first energy eigenvalue of a
neutron trapped in a box of width 15 $\mu$m is $E_1^{\mathrm{Box}}
= 0.9 \ \mathrm{peV}$, of the same order of magnitude as the first
energy eigenvalue of a neutron in the Earth's gravitational field,
$E_{1} = 1.4 \ \mathrm{peV}$. For more realistic potentials it is
possible to reproduce the first energy level of $E_1 = 1.4 \
\mathrm{peV}$ at an opening of 15 $\mu$m, explaining the
``gravitational quantum energy state" as merely a normal geometric
cavity-effect\footnote{If the transverse neutron temperature is
$20 \ \mathrm{nK}$ as stated in \cite{Schwarz}, corresponding to
$\sim 1 \ \mathrm{peV}$, even the simple infinite box potential
can explain the first step. The smallest separation ($a \simeq 15
\ \mu$m) then corresponds to the high energy ``tail" of the
transverse neutron energy. Any free-falling neutrons with higher
transverse energies, which could traverse narrower slits, are
``filtered out" by the absorber arrangement, just as in the
original argument \cite{Nev,Nev1}.}. A thorough investigation
would necessitate a very exact and complicated modelling of the
potential at the mirror and, especially, at the absorber.

However, as the final deciding factor in physics is
\textit{experiment} there is, at least in principle, a much
simpler way to check this. Keeping everything else identical, turn
the cavity from being horizontal to being vertical! If the effect
is due to gravity it must then disappear as a potential well in
the vertical direction no longer is present. If the effect is
sustained in the vertical configuration it is solely due to the
geometry of the cavity and the intrinsic properties of the neutron
beam. (An even more ideal way would be to do the measurements in
free-fall, but this seems virtually impossible to do in practice.)
Nesvizhevsky \textit{et al.}, have controlled this by ``reversing
the geometry", \textit{i.e.}, placing the absorber at the bottom
instead of above. This control, however, is inconclusive. The
absorber length is 13 cm, the mirror length is 10 cm. Outside the
cavity formed between mirror and absorber the (unquantized)
neutrons fall freely. Inside the cavity, even in the ``reversed"
case due to the fact that the absorber is non-ideal, there is a
\textit{standing neutron-wave}, meaning that the neutrons do not
fall at all. When using the standard configuration the absorber is
at the top, unable to absorb down-falling neutrons (outside the
cavity). In the reversed configuration, however, there is a 3 cm
``excess" of absorber outside, and below, the mirror, drastically
reducing the neutron flux into the detector (as observed). Because
of this, one can unfortunately not rule out a purely geometrical
explanation of the measured effect, based on the performed tests.
\section{Conclusions and suggestions for improvements}
Our conclusion is that a quantization of the gravitational ground
state of a neutron has not been unambiguously identified. A
``normal" cavity-effect can also explain the first discrete
increase in the neutron count $N$, which is the only experimental
result claimed to underlie the identification.

We therefore propose the following improvements of the experiment:
\begin{itemize}
  \item Rotating the experimental setup by $90^\circ$ keeping
  everything else, and especially the transverse neutron energies,
  constant. This gives a vertical instead of horizontal cavity.
  If the same result still occurs this would indicate that it
  is due only to the geometry of the experimental setup, as
  no gravitational quantum states can form in this case.
  \item Measuring \textit{where} the neutrons strike the detector.
  This should be possible in principle, although not yet in
  practice.
  Since there is a standing neutron wave, the neutron is not falling
  in a classical sense, and the observed probability
  distribution should reflect $|\psi|^2$. This would directly discriminate
  between different theoretical explanations, as the linear gravitational potential
  gives a very distinct probability distribution, different from those arising
  from ``cavity-potentials". However, it would require
  moving the (improved) detector right up to the end of the cavity, with no
  ``free" mirror surface as in the present setup. As this test is not
  currently possible, we instead propose that measurements should be made of
  $N = N(x)$ as a function of the cavity length, $x$, accordingly varied.
  When the neutron wave function penetrates the absorber, the
  neutron count in the detector should be $N \propto e^{-k x}$,
  where $k = k(\Delta h)$ is proportional to the fraction of $|\psi|^2$ inside
  the absorber\footnote{A constant absorption probability per unit
  length, $dN = - N \, k \, dx$, gives $N = N_0 \, e^{-k  x}$.}.
  In this way it should be possible to differentiate between theoretical explanations
  (gravitation/cavity potentials) of the neutron counts.
\end{itemize}
The experimental group will also try to measure transitions
between different quantum states. We close by noting that if this
actually is accomplished it would be the first (indirect)
measurement of a graviton spectrum, analogous to normal
electromagnetic photon spectra from atoms. As one-graviton
exchange is overwhelmingly more likely than multi-graviton
exchange (suppressed by powers of the very small gravitational
coupling constant) the transition energy difference, $\Delta E$,
will be carried away as a graviton with wavelength $\lambda_{grav}
= h c /\Delta E \sim 10^6$ m.

\end{document}